# Intermediate Performance of Rateless Codes

Sujay Sanghavi
LIDS, MIT
sanghavi@mit.edu

*Abstract*— Rateless/fountain codes are designed so that all input symbols can be recovered from a slightly larger number of coded symbols, with high probability using an iterative decoder. In this paper we investigate the number of input symbols that can be recovered by the same decoder, but when the number of coded symbols available is *less than* the total number of input symbols. Of course recovery of all inputs is not possible, and the fraction that can be recovered will depend on the output degree distribution of the code.

In this paper we *(a)* outer bound the fraction of inputs that can be recovered for *any* output degree distribution of the code, and *(b)* design degree distributions which meet/perform close to this bound. Our results are of interest for real-time systems using rateless codes, and for Raptor-type two-stage designs.

## I. INTRODUCTION

Rateless codes [1, 2], or fountain codes, are random linear codes designed for communication over erasure channels. Given a block of input symbols, a rateless encoder generates a potentially infinite sequence of output symbols, each of which is generated identically and independently. This process of generation is designed so that it is possible to recover all input symbols from any sligtly larger set of output symbols, with high probability when the total number of intput symbols is large enough. This implies that rateless codes are universally capacity achieving. Further, this recovery property can be achieved via a simple iterative decoder of low complexity.

The construction of rateless codes makes them appealing for myriad applications. In general, rateless codes are seen to perform well for scenarios where the erasure probability/pattern is not known, and in multicast/broadcast applications where the encoder outputs onto a shared medium and cannot tune its transmissions to individual receivers.

The design of rateless codes has been optimized so that the low-complexity decoder can recover all inputs provided it starts with slightly more outputs than inputs. In this paper, we investigate the *intermediate performance*, i.e. the case when the number of received output symbols is *less than* the number of input symbols. Of course in this case it is not possible to fully recover all input symbols. We investigate the fraction of input symbols that can be recovered – *by the same iterative decoder* – as a function of the number of received output symbols and the randomized method by which the codes are generated.

The motivations for this investigation are two fold:

- Codes can be designed – as done in [2] – so that it is sufficient to decode only a large enough fraction of the inputs, instead of all inputs. The implications of our results for the design of such codes are discussed in Section V.
- There may be real-world scenarios where users do not receive sufficient output symbols, and scenarios where it is sufficient to recover a large enough fraction of the inputs instead of all inputs.

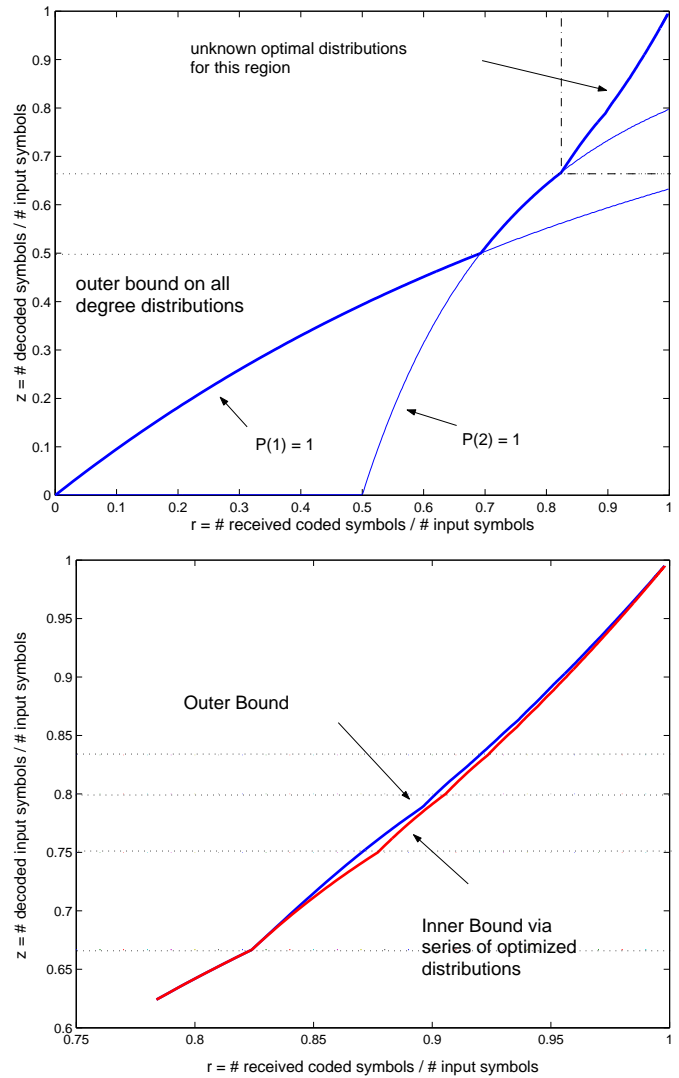

Fig. 1. **Depiction of main results.** Each figure plots the (asymptotic) fraction $z$ of recoverable inputs as a function of the normalized number $r$ of received outputs. The bold line in the top figure represents a pointwise outer (i.e. upper) bound on $z$ as a function of $r$, for *any* output degree distribution used to generate the rateless code. This bound is tight for $z \in [0, \frac{2}{3}]$ and $r \in [0, \frac{3}{4} \log 3]$, and the corresponding optimal degree distributions are marked. Beyond this region optimal distributions are not known. The second figure is this unknown region of the first figure expanded, with the outer bound as marked. For this region we develop a series of degree distributions $P_r$, one for each $r$. The "inner bound" represents the points $(r, z_r)$ where $z_r$ is the corresponding $z$ achieved by $P_r$ at $r$. It can be seen that our distributions perform close to (but below) the outer bound.

As an example, consider a scenario where a data stream is to be transmitted to multiple users over a shared medium. The stream needs to be decoded in real time, or near-real time with a finite amount of buffering. If rateless codes are to be used for the transmission, the stream would have to be broken into blocks of symbols, and each block would have to be encoded seperately. These output symbols would then be transmitted over the shared medium. If such a transmission strategy is employed, the output symbols from any given block of inputs will only be transmitted for a finite amount of time, before the encoder moves on to the next block of inputs.

If the user channel qualities are very disparate, it is possible that some user may not receive the requisite number of outputs for some input block. It is reasonable to assume that many real-time applications, if suitably pre-coded, can be reliably played back from a large enough fraction of the inputs. Thus it is of interest to understand the intermediate performance – how many inputs can be recovered by the decoder from an insufficient number of output symbols.

As it turns out, any rateless code designed to achieve capacity will have poor intermediate performance: even if the number of outputs is very close to (but not exactly) sufficient, the fraction of inputs that can be decoded will be negligible. This fragile performance of capacity achieving rateless codes motivates us to investigate other rateless codes. In particular, each rateless code is associated with a *degree distribution*, and we investigate and bound the performance of all distributions, not just the capacity-achieving ones. Figure 1 describes the main results of this paper.

In the following, we first give a brief background on rateless codes in Section II and then develop our investigation of intermediate performance in Section III. The main results are derived in Section IV, followed by a discussion in Section V.

## II. BACKGROUND: RATELESS CODES

In this section we briefly describe how rateless codes are encoded and decoded. A more thorough explanation and treatment can be found in [1]. Throughout this paper, we will assume for simplicity that the input and output symbols are bits[1]

**Encoding:** Given $k$ input symbols and a probability distribution $P$ over $[k]$, encoder generates each output as follows: *(a)* it chooses a random degree $d$ according to $P$, and *(b)* chooses a set of $d$ inputs uniformly at random and XORs them to get the output.

**Decoding:** Given $n$ output symbols, the decoder selects one of degree one. The value of the corresponding input is set, and that input is then cancelled out of all other outputs it is a part of. Decoding stops when there are no degree-ones left.

The degree distribution is a crucial component of the rateless code design above. Clearly, for any code, the decoder needs $n \geq k$ to recover all input bits. LT Codes [1], the first rateless codes, were designed with the objective of ensuring that all $k$ input bits are recovered with high probability from $n$ bits when is not much larger than $k$. In [1] an *ideal soliton* distribution $I_k$ is identified as being the unique distribution that will, in expectation, enable efficient decoding with the iterative decoder. This distribution performs poorly due to the random fluctuations, and is thus modified into a *robust soliton* distribution $P_k^{(R)}$ that works well in practice. In the limit as $k \to \infty$, both $I_k$ and $P_k^{(R)}$ converge pointwise to the following distribution

$$I(i) = \begin{cases} 0 & \text{if } i = 1 \\ \frac{1}{i(i-1)} & \text{if } i \geq 2 \end{cases} \quad (1)$$

This $I$ is unique: any sequence of degree distributions that achieve capacity with iterative decoding will have $I$ as the limiting distribution.

Recall that the iterative decoder requires degree-one outputs to start decoding. While for each finite $k$ both $P_k^{(R)}$ and $I_k$ will result in some degree-one parities being generated, the above limiting distribution has no degree-one output parities. This thus illustrates that given a limiting degree distribution with some predicted prformance, it is possible to modify it for finite $k$ so as to actually achieve the predicted performance.

## III. PROBLEM STATEMENT: INTERMEDIATE PERFORMANCE

In general, the (distribution of) the number of input bits recovered will depend on the number of received parities, the degree distribution used for generating the rateless codes, and the total number $k$ of input bits. In this paper we will be concerned with asymptotic performance, i.e. we for the case when $k$ is large. In particular, given a sequence $\{P_k\}$ of degree distributions that converge pointwise to a distribution $P$, we will be concerned only with the performance of $P$. Towards this end we define the following two quantities

$$r_k = \frac{\text{no. of received parity bits}}{\text{no. of input bits } k}$$
$$z_k = \frac{\text{no. of decoded input bits}}{\text{no. of input bits } k}$$

We will be interested in the relation between $r$ and $z$ as $k \to \infty$. Towards this end we define $z(r, P)$ to be the limiting value of $z_k$ for $r_k = r$ and $P_k \to P$. [2]

As a simple illustrative example, consider again the soliton distribution of [1]. In terms of the notation in this paper, the result of [1] implies that $z(1, I) = 1$. Note that this is the same as saying that $I$ is a capacity achieving distribution.

As pointed out in [4], recent results in random hypergraphs by Darling and Norris [3] can be used to study the asymptotic performance of the iterative decoder for the limiting degree

---

[1] In practice each symbol is a packet – a vector of bits – and the operations described below are carried out in parallel for each position of the vector.

[2] [3] shows that these limits are well defined.

distribution of the rateless code. To do so we will need some additional notation.

First, for any degree distribution $P$, we define its generating function $\mathcal{P}(t) = \sum_{i \geq 1} P(i) t^i$ and its derivative $\mathcal{P}'(t)$. Further, for any real number $r > 0$ we define the term

$$s(r, P) = \inf \{t \in [0,1) : r\mathcal{P}'(t) + \log(1-t) < 0\} \wedge 1 \quad (2)$$

We now restate Theorem 2.1 of [3] in the notation developed in our paper.

*Theorem 1 (Darling and Norris [3]):* Let real number $r > 0$ and the limit degree distribution $P$ be such that

$$r\mathcal{P}'(t) + \log(1-t) > 0 \quad \text{for } 0 \leq t < s(r, P) \quad (3)$$

Then, as $k \to \infty$, if the number of received parity bits is $Poisson(rk)$ then $z_k \to s(r, P)$.

The above theorem gives a way to calculate the quantity $z(r, P)$ of interest : modulo the poisson approximation, it says that for limiting distriulos that satisfy the conditions of the theorem we have that $z(r, P) = s(r, P)$.

Note that for the ideal soliton ditribution $I$ we have that $s(1, I) = 1$. However the above theorem, as stated, does *not* apply to the limiting soliton distribution $I$. This is because $\mathcal{I}'(0) = 0$ and so for any $r$ we will violate the condition $r\mathcal{I}'(0) + \log(1 - 0) > 0$. In fact, for $r = 1$ we have that $\mathcal{I}'(t) + \log(1-t) = 0$ for all $0 \leq t \leq 1$. Thus the above theorem cannot be directly used to show that $z(1, I) = 1$.

The above problem with the soliton distribution illustrates why it may be hard to use the above theorem directly to evaluate the performance of otherwise interesting degree distributions. As a stepping stone to obtaining interesting results using the above ideas, we first define a perturbation as follows: given a sequence of degree distributions $P_k \to P$ and a real number $\delta > 0$, let $Q_k$ be the distribution defined by

$$Q_k(1) = \delta + (1-\delta) P_k(1)$$
$$Q_k(i) = (1-\delta) P_k(i) \quad \text{for } i \geq 2$$

In the above we have moved a small amount of mass from the higher degrees of $P$ to degree one. Note that now $Q_k \to Q$ whose generating function is $\mathcal{Q}(t) = (1-\delta)\mathcal{P}(t) + \delta t$. We now use the above theorem to show that the performance of the perturbed $Q$ will be close to that predicted by $s(r, P)$, even if the original $P$ does not satisfy the conditions (3).

*Lemma 1:* Given $\epsilon > 0$ there exists $\delta_1 > 0$ so that for and $\delta < \delta_1$ the corresponding perturbed $Q$, as defined above, the following holds

$$s(r, P) \leq z(\frac{r}{1-\delta}, Q) \leq s(r, P) + \epsilon$$

*Remark:* The above lemma says that, with a slightly higher value of $r$, the $\delta$-perturbed distribution has the actual fraction $z$ of recovered inputs close to the $s(r, P)$ predicted for $P$ and $r$.

*Proof:* Note that $\frac{r}{1-\delta}\mathcal{Q}'(t) = r\mathcal{P}'(t) + \frac{\delta}{1-\delta}$ from which it follows that $s(\frac{r}{1-\delta}, Q) \geq s(r, P)$. It also follows that, given $\epsilon$, there exists a $\delta_2$ such that $\delta < \delta_2$ implies that $s(\frac{r}{1-\delta}, Q) \leq s(r, P) + \epsilon$.

Also, there exists $\delta_3$ so that for all $\delta < \delta_3$ we have that property (3) is satisfied by $\frac{r}{1-\delta}$ and $Q$. For any such $\delta$ we have that $z(\frac{r}{1-\delta}, Q) = s(\frac{r}{1-\delta}, Q)$ from Theorem 1. Thus setting $\delta_1 = \min\{\delta_2, \delta_3\}$ proves the lemma. ∎

In light of the above lemma, for the remaining portion of the paper we will be interested studying and bounding $s(r, P)$ as the degree distribution $P$ is allowed to vary.

## IV. RESULTS

The quantity $s(r, P)$ exists for all $P$, and hence we can easily find out the (approximate, asymptotic) relation between the number of decoded inputs and the number of received parities for any particular rateless code degree distribution $P$. In this section we take this as the starting point in our investigation of the intermediate performance of rateless codes. Lemma 1 in the last section justifies this approach.

As an illustrative first step, we investigate the performance of the limiting soliton distribution $I$ specified by (1). The corresponding generating function is $\mathcal{I}'(t) = -\log(1-t)$. It is easy to see that there is a discontinuity at $r = 1$: $s(1, I) = 1$ but $s(r, I) = 0$ for every $0 \leq r < 1$. This means that while the soliton achieves capacity, its performance is fragile in the sense that even if the number of received parities is slightly less than required the fration of recovered inputs will be very small.

We are interested in distirbutions with optimal intermediate performance. Towards this end we define the following terms: for each $0 \leq r < 1$ let $P_{(r)}$ be s.t. $s(r, P_{(r)}) \geq s(r, P)$ for all $P$. We will be interested in finding, for each $r$, the distribution $P_{(r)}$ and corresponding value $s(r, P_{(r)})$. For the purpose of analysis, it is convenient to define for each $0 \leq z < 1$ the equivalent terms

$$r(z, P) = \inf\{r : s(r, P) \geq z\}$$
$$P_{(z)} \quad \text{s.t.} \quad r(z, P_{(z)}) \leq r(z, P) \quad \text{for all } P$$

$P_{(z)}$ is the optimal distribution for $z$: it will enable the decoding of a fraction $z$ of the inputs from the smallest number of received output symbols. Characterizing/bounding $P_{(z)}$ and corresponding value $r(z, P_{(z)})$ for each $z$ is equivalent to a characterization in terms of $r$, and we will find this more convenient.

The above objective is achieved exactly for $z \in [0, \frac{2}{3}]$. For the remaining values we do not know an exact characterization, so we do two things for each $z \in (\frac{2}{3}, 1)$:
1) Use linear programming duality to lower (i.e. outer) bound the $r(z, P_{(z)})$.
2) Design distirbutions $\widehat{P}_{(z)}$ that perform close to this outer bound.

Note that a lower bound of $r(z, P_{(z)})$ represents an outer bound, i.e. it is a quantity that may not be achievable by any rateless code degree distribution.

In the task of finding $P_{(z)}$ the following lemma provides an important simplification.

*Lemma 2:* Given $z < 1$, if integer $m \geq 1$ is such that $z < \frac{m}{m+1}$ then it has to be that $P_{(z)}(i) = 0$ for all $i \geq m+1$.

Alternatively, if $m$ is such that $z = \frac{m}{m+1}$ then there exists an optimal $P_{(z)}$ with $P_{(z)}(m+1) = 0$.

*Proof:*

Recall that $\mathcal{P}'(t) = \sum_{i \geq 1} P(i) \, i t^{i-1}$. Now, if $t \leq \frac{m}{m+1}$ then for every $n \geq m+1$ we have that $(n-1)t^{n-2} > nt^{n-1}$. Thus, in particular, we have that $mt^{m-1} > nt^{n-1}$.

Now, suppose $P_{(z)}$ is such that $\sum_{i>m} P_{(z)}(i) > 0$. Then construct a new $\widetilde{P}_{(z)}$ as follows:

$$\widetilde{P}_{(z)}(i) = P_{(z)}(i) \text{ for } 1 \leq i \leq m-1$$
$$\widetilde{P}_{(z)}(m) = \sum_{i \geq m} P_{(z)}(i)$$

Then, it follows that the corresponding generating functions will satisfy $\widetilde{\mathcal{P}}'_{(z)}(t) > \mathcal{P}'_{(z)}(t)$ for all $0 < t < \frac{m}{m+1}$ and $\widetilde{\mathcal{P}}'_{(z)}(t) \geq \mathcal{P}'_{(z)}(t)$ for $t = \frac{m}{m+1}$.

For $z < \frac{m}{m+1}$ this means that $r(z, \widetilde{P}_{(z)}) < r(z, P_{(z)})$, contradicting the choice of $P_{(z)}$. Thus for such a $z$ it has to be that $\sum_{i>m} P_{(z)}(i) = 0$.

For $z = \frac{m}{m+1}$ the above means that $r(z, \widetilde{P}_{(z)}) \geq r(z, P_{(z)})$ and since $P_{(z)}$ is optimal this will be equality. This means that the alternate distribution $\widetilde{P}_{(z)}$ is also optimal for $z$. The lemma is thus proved. ∎

As an immediate corollary of the above lemma we have that for all $0 \leq z \leq \frac{1}{2}$ the optimal distribution is $P_{(z)}(1) = 1$ and the corresponding $r(z, P_{(z)}) = -\log(1-z)$ for $0 \leq z \leq \frac{1}{2}$. In terms of $r$, this means that $P_{(r)}(1) = 1$ and $z(r, P_{(r)}) = 1 - e^{-r}$ for $0 \leq r \leq \log 2$.[3]

Moving on to higher values of $z$ and $r$, Lemma 2 means that for any $z < 1$ if $m$ is such that $\frac{m-1}{m} \leq z \leq \frac{m}{m+1}$ then without loss of generality we can restrict attention to degree distributions that have support on $[m]$. However, it does not immediately provide an exact answer for what the optimal distributions $P_{(r)}$ or $P_{(z)}$ are. So we now use linear programming duality to provide a bound on $r(z, P_{(z)})$.

Given a fixed $z < 1$, the optimal $r(z, P_{(z)})$ is the solution of the following optimization problem:

$$\min_{r,P} r \quad \text{s.t.}$$
$$r\mathcal{P}'(t) + \log(1-t) \geq 0 \quad \text{for } 0 \leq t < z$$

The problem is stated above is not linear since the constraints are not linear. However, it can be easily converted into a linear program by a change of variables. Specifically, let sequence $a$ be defined by $a(i) = rP(i)$ and its generating function be $\mathcal{A}(t) = \sum_i a(i) t^i$. Clearly $r = \sum_i a(i)$ and $r\mathcal{P}(t) = \mathcal{A}(t)$. If integer $m$ is such that $\frac{m-1}{m} \leq z \leq \frac{m}{m+1}$ then the above problem can then be rewritten so that $r(z, P_{(z)})$ is the solution of

$$\min_a a(1) + \ldots + a(m) \quad \text{s.t.}$$
$$\mathcal{A}'(t) + \log(1-t) \geq 0 \quad \text{for } 0 \leq t < z \quad (4)$$

---

[3]Thus means that if the objective of communication is only the recovery of at least half of the inputs, then it is optimal to not employ any coding at all !

This optimization problem is now linear, but with infinite dimensional constraints. Nevertheless, it falls within the standard theory of such linear programs. Clearly, the problem is feasible. Also, we can write down the dual problem

$$\min_{f(x)} E[-\log(1-X)] \quad \text{s.t.}$$
$$X \in [0, z] \text{ and } E[X^{i-1}] \leq \frac{1}{i} \text{ for } 1 \leq i \leq m \quad (5)$$

where the optimization has to be carried out over all distributions $f(x)$ of the random variable $X$ that have support only in $[0, z]$.

The dual linear programs above will have no duality gap. Thus, we can prove optimality of a particular candidate distribution by evaluating the corresponding value of the primal objective function, and the constructing a dual solution of equal value. This allows us to calculate the optimal distribution $P_{(z)}$ and corresponding value $r(z, P_{(z)})$ for a further range of $z$'s.

*Lemma 3:* For $z \in [\frac{1}{2}, \frac{2}{3}]$, the optimal distribution is given by $P_{(z)}(2) = 1$, and the corresponding $r(z, P_{(z)}) = \frac{-\log(1-z)}{2z}$.

*Proof:*

Consider the solution $a(2) = \frac{-\log(1-z)}{2z}$ and $a(i) = 0$ for all $i \neq 2$ for the primal. This satisfies the constraint: $\mathcal{A}'(0) = 0$ and

$$\mathcal{A}'(t) + \log(1-t) = 2t \left( \frac{-\log(1-z)}{2z} + \frac{\log(1-t)}{2t} \right)$$
$$\geq 0 \quad \text{for } 0 < t < z$$

where the last inequality follows from the fact that $\frac{-\log(1-t)}{2t}$ is a strictly increasing function of $t$. Thus the above solution is valid for the primal problem and yields a value of $\frac{-\log(1-z)}{2z}$.

Consider the dual solution

$$f(x) = \left(1 - \frac{1}{2z}\right)\delta(x) + \frac{1}{2z}\delta(x-z)$$

which puts mass $\frac{1}{2z}$ on the point $z$ and the remaining mass at the origin. This satisfies the constraints because $E[X] = z(\frac{1}{2z}) = \frac{1}{2}$. Thus it is valid solution, and yields a value of $\frac{-\log(1-z)}{2z}$ for the dual. This is seen to be equal to the value of our guess for the primal, and thus by weak duality we have that both are optimal and that $r(z, P_{(z)}) = \frac{-\log(1-z)}{2z}$. ∎

Restated in terms of $r$, the above lemma says that for $\log 2 \leq r \leq \frac{3}{4} \log 3$ we have that the optimal distribution is $P_{(r)}(2) = 1$.

Unfortunately, we have not been able to extend the above ideas of linear programming duality to exactly characerize $P_{(z)}$ or $r(z, P_{(z)})$ for values of $z > \frac{2}{3}$. Nevertheless, we can use linear programming to provide outer bounds.

Specifically, note that the value of any feasible solution of the dual problem (5) will be a lower bound of the value of the primal problem, and thus of $r(z, P_{(z)})$. The dual problem can be approximately solved numerically by assuming that the optimal $f(x)$ has support only on a uniform grid of fine granularity over $t \in [0, z]$. Under this assumption the dual problem becomes a standard linear program with finite numbers of variables and constraints. Its value will be a true lower bound on $r(z, P_{(z)})$. This outer bound with a grid spacing of 0.001 is shown in Figure 1.

The outer bounds as computed above serve as a benchmark to evaluate the merit of any particular candidate distribution $P$ in the range $z \in (\frac{2}{3}, 1)$ where the corresponding exact $P_{(z)}$ is not known. In the following we develop, for each $z \in (\frac{2}{3}, 1)$, a corresponding $\widehat{P}_{(z)}$. The performance of these distributions is plotted in Figure 1, and we see that it is close to that of the outer bound.

The intuition behind the design is as follows. Recall that with the optimal $P_{(z)}$ for $z \in [0, \frac{2}{3}]$, the constraint of the primal problem (4) is tight only at points $t = 0$ and $t = z$. [4] However, for $z \in (\frac{2}{3}, 1)$ the constraints may be tight at other points besides 0 and $z$.[5] Nevertheless, we can design the $\widehat{P}_{(z)}$ so that the constraints of the primal (4) are tight only at $t = 0$ and $t = z$. We do so in the following lemma.

*Lemma 4:* Given a fixed $z \in (\frac{2}{3}, 1)$, let integer $m$ be such that $\frac{m-1}{m} \leq z \leq \frac{m}{m+1}$, and let $\widehat{P}_{(z)}$ be defined by

$$\widehat{P}_{(z)}(i) = \frac{1}{ai(i-1)} \quad \text{for } 2 \leq i \leq m-1$$
$$\widehat{P}_{(z)}(m) = 1 - \frac{m-2}{a(m-1)}$$
$$\widehat{P}_{(z)}(i) = 0 \quad \text{for all other } i$$

where

$$a = \frac{m-1}{m} + \frac{1}{mz^{m-1}} \sum_{i \geq m} \frac{z^i}{i} \quad (6)$$

This $\widehat{P}_{(z)}$ represents a probability distribution, and we have that

$$a\widehat{\mathcal{P}}_{(z)}(t) + \log(1-t) > 0 \quad \text{for } 0 < t < z$$

So the asymptotic fraction of decoded inputs will satisfy $s(a, \widehat{P}_{(z)}) = z$. This is the same as saying $r(z, \widehat{P}_{(z)}) = a$.

*Proof:*
We first verify that the $\widehat{P}_{(z)}$ is a probability distribution. Note that $\widehat{P}_{(z)}(i) \geq 0$ for all $a \geq \frac{m-2}{m-1}$, and further that

$$a \sum_i \widehat{P}_{(z)}(i) = \sum_{i=2}^{m-1} \frac{1}{i(i-1)} + a - \frac{m-2}{m-1} = a$$

and thus $\sum_i \widehat{P}_{(z)}(i) = 1$. So it is a probability distribution for any $a \geq \frac{m-2}{m-1}$. We now prove the second property. It is easy to verify that

$$a\widehat{\mathcal{P}}_{(z)}(t) + \log(1-t) = amt^{m-1} - (m-1)t^{m-1} - \sum_{i \geq m} \frac{t^i}{i}$$

---

[4] In terms of the dual (5), this means that the optimal $f(x)$ will have support only on the points $x = 0$ and $x = z$.

[5] Indeed, numerical simulations seem to indicate that there are always a finite number of support points, but that this number increases as $z \to 1$. This is consistent with the fact that for $z = 1$ the optimal soliton distribution $I$ will result in the optimal dual having support on the entire $[0, 1]$ interval.

So, $a\widehat{\mathcal{P}}_{(z)}(t) + \log(1-t) > 0$ if and only if $t > 0$ and

$$a > \frac{m-1}{m} + \frac{1}{mt^{m-1}} \sum_{i \geq m} \frac{t^i}{i}$$

Note that in the above expression the RHS is a strictly increasing function of $t$, and that the value of $a$ – (given in (6) – is exactly the RHS of the above expression at $t = z$. Hence the above expression is true, proving the lemma. The fact that $s(a, \widehat{P}_{(z)}) = z$ follows from the above and the definition (2). ■

In light of the above lemma, we can now see for each $z$ how the corresponding $r(z, \widehat{P}_{(z)}) = a$ as defined by (6) compares to the numerically computed outer bound for that $z$ obtained by discretizing the dual. This comparison is made in the second subfigure of Figure 1.

## V. Discussion

Note that the approximate distributions $\widehat{P}_{(z)})$ presented in Lemma 4 are pretty close to the ideal soliton $I$. In particular, as $z \to 1$, $\widehat{P}_{(z)}) \to I$. It can thus be thought of as an approrpiate truncation and rescaling of the ideal soliton $I$.

A different truncation and rescaling of $I$ is given in the paper on Raptor codes [2, Sec. 6]: for $\epsilon > 0$ they define $D = \lceil 4(1+\epsilon)/\epsilon \rceil$ and $\mu = \epsilon/2 + (\epsilon/2)^2$ and a probability distribution whose generating function is

$$\Omega_D(t) = \frac{1}{1+\mu}\left(\mu t + \sum_{i=2}^{D} \frac{t^i}{i(i-1)} + \frac{t^{D+1}}{D}\right)$$

It is shown that this distribution can recover $z = 1 - \delta$ from $r = 1 + \epsilon$, and thus can get withing $\epsilon$ of capacity. Our results suggest that using the distribution $\widehat{P}_{(1-\delta)}$ as developed in 4 will give better performance, because it will enable recovery of $z = 1 - \delta$ with $r < 1$ as opposed to $r = 1 + \epsilon$ as is the case in [2].

While the above distinction is small for small $\epsilon$ and $\delta$, it will be significant for moderate values. This may be of interest if the pre-code in [2] is of small length and not too close to capacity achieving, and thus requires a larger $\delta$.

Solving for the exact optimal distribution in the $z \in (\frac{2}{3}, 1)$ region remains of interest.

## Acknowledgment

The author would like to acknowledge helpful discussions with Bruce Hajek.